\documentclass[iop]{aastex62}

\usepackage{hhline}
\usepackage{natbib}
\usepackage{graphicx}
\usepackage{subfigure}
\usepackage{amsmath}
\usepackage[utf8]{inputenc}
\usepackage{float}
\usepackage{gensymb}
\usepackage [english]{babel}
\usepackage [autostyle, english = american]{csquotes}
\MakeOuterQuote{"}
\usepackage{multirow}
\usepackage{array, xcolor}
\newcolumntype{M}[1]{>{\centering\arraybackslash}m{#1}}
\newcolumntype{N}{@{}m{0pt}@{}}
\usepackage{soul}

\shortauthors{Tabeshian et al.}

\begin{document}

\title{Asteroid (3200) Phaethon: colors, phase curve, limits on cometary activity and fragmentation}

\author{Maryam Tabeshian}
\affiliation{Department of Physics and Astronomy, The University of Western Ontario, London, ON, N6A 3K7, Canada}
\author{Paul Wiegert}
\affiliation{Department of Physics and Astronomy, The University of Western Ontario, London, ON, N6A 3K7, Canada}
\affiliation{Centre for Planetary Science and Exploration, The University of Western Ontario, London, ON, N6A 3K7, Canada}
\author{Quanzhi Ye}
\affiliation{Division of Physics, Mathematics and Astronomy, California Institute of Technology, Pasadena, CA, 91125, USA}
\affiliation{Infrared Processing and Analysis Center, California Institute of Technology, Pasadena, CA, 91125, USA}
\author{Man-To Hui}
\affiliation{Department of Earth, Planetary and Space Sciences, UCLA, Los Angeles, CA, 90095, USA}
\author{Xing Gao}
\affiliation{Xinjiang Astronomical Observatory, Urumqi, Xinjiang 830011, China (mainland)}
\author{Hanjie Tan}
\affiliation{Department of Physics, National Central University, 32001, Taiwan}
\textit{Accepted by the Astronomical Journal, 23 May 2019}

\email{Corresponding author email: mtabeshi@uwo.ca}


\begin{abstract}
We report on a multi-observatory campaign to examine asteroid 3200
Phaethon during its December 2017 close approach to Earth, in order to
improve our measurements of its fundamental parameters, and to search
for surface variations, cometary activity and fragmentation. The mean
colors of Phaethon are B-V = $0.702\pm0.004$, V-R = $0.309\pm0.003$,
R-I = $0.266\pm0.004$, neutral to slightly blue, consistent with
previous classifications of Phaethon as a F-type or B-type asteroid
\citep{Tholen85, Green95}. Variations in Phaethon's B-V colors (but
not V-R or R-I) with observer sub-latitude are seen and may be
associated with craters observed by the Arecibo radar
\citep{Taylor19}. High cadence photometry over phases from 20 to 100
degrees allows a fit to the values of the HG photometric parameters;
$H = 14.57 \pm 0.02, 13.63 \pm 0.02, 13.28 \pm 0.02, 13.07\pm 0.02$;
$G = 0.00 \pm 0.01, -0.09 \pm 0.01, -0.10 \pm 0.01, -0.08 \pm 0.01$ in
the BVRI filters respectively; the negative G values are consistent
with other observations of F type asteroids e.g \citep{Lagerkvist90}.
Light curve variations were seen that are also consistent with
concavities reported by Arecibo, indicative of large craters on
Phaethon's surface whose ejecta may be the source of the Geminid
meteoroid stream. A search for gas/dust production set an upper limit
of $0.06\pm 0.02$ kg/s when Phaethon was 1.449 AU from the Sun, and
$0.2\pm0.1$ kg/s at 1.067 AU. A search for meter-class fragments
accompanying Phaethon did not find any whose on-sky motion was not
also consistent with background main belt asteroids.

\end{abstract}



\section{Introduction}
\label{Sec:Intro}

Although dynamically associated with the Geminid meteoroid stream \citep{Whipple83}, the absence of cometary activity or mass loss from the near-Earth asteroid (3200) Phaethon for over the two decades since its discovery has made it an interesting target for numerous studies. Given the short dynamical lifetime of Geminid meteoroids ($\sim 10^3$~yr; \cite{Ryabova07}), astronomers speculated that Phaethon must have undergone some recent cometary activity which may have continued to the present time. This idea was further strengthened when NASA-STEREO coronal imaging observations of Phaethon in 2009 by \cite{Jewitt10} and three years later by \cite{Li13} showed anomalous and sudden brightening of the asteroid at perihelion ($q = 0.14$ AU) by a factor of 2 which was attributed to the release of solid grains from the nucleus. Further observations using the NASA-STEREO coronal imaging spacecraft by \cite{Jewitt13} confirmed that Phaethon had a comet-like dust tail when it was just past perihelion in both 2009 and 2012; \cite{Hui17} analyzed Phaethon around its 2016 perihelion and observed similar behavior. However, whether Phaethon is an asteroid or an extinct cometary nucleus is still a matter of debate.  

While \cite{Davies86} had suggested that Phaethon is an extinct comet, \cite{Jewitt10} argued the dust grains were likely released through desiccation cracking of the surface and thermal fracture due to extensive heating ($\sim 1000 ~K$) near perihelion rather than volatile sublimation. From the properties and morphology of the tail, the effective radius of the dust particles and their combined mass were estimated to be $\sim 1 ~\mu$ m and $3\times10^5$ kg, respectively, too low to explain the current Geminid meteoroid stream. On the other hand, using a 3D model of gas and heat transport in porous sub-surface layers of Phaethon's interior, \cite{Boice13} found that relatively pristine volatiles in the interior of Phaethon might still exist despite many perihelion passages over its short (1.49 yr) orbital period. Thus high-temperature processes as well cometary outgassing may both be at work and the exact mechanism or mechanisms by which the Geminid meteoroid stream was produced remain unclear.

Even close to perihelion where Phaethon shows a dust tail, not only does it produce 100 to 1000 times less mass than needed, individual dust particles that are ejected from Phaethon are much smaller and less massive than typical meteoroids in the Geminid stream and are quickly removed by solar radiation pressure \citep{Jewitt10}. Several alternative mechanisms have been proposed to explain how Phaethon could supply the Geminid stream mass. For instance, according to \cite{Jewitt18}, it is possible that the Geminids were produced as a result of a catastrophic event of unknown origin that occurred at some point within the past few thousand years. On the other hand, \cite{Yu18} considered the possibility that Phaethon originated beyond the ice line - possibly having broken away from asteroid 2 Pallas as Phaethon's reflectance spectra shows striking similarities to that of Pallas \citep[see][]{deLeon10}. Less than 1 Myr ago Phaethon would have moved to its present orbit, and so could have retained its sub-surface ice until the present day.

Here we report ground-based observations of Phaethon by Gemini and the Canada-France-Hawaii Telescope (CFHT) on Maunakea, Hawaii as well the Xingming Observatory on Mt. Nanshan, Xinjiang, China obtained in November-December 2017 as the asteroid was approaching Earth. Our primary goals were to establish its colors and phase curve, to look for photometric variations across Phaethon's surface, to establish whether it has a coma and measure any mass loss during the observation period, and to search for meter-class fragments which might indicate a different mass-loss mechanism. We start in Section \ref{Sec:Obs} by describing our observations, and then move to our analysis in Section \ref{Sec:Results}. Finally, we present a summary and conclusions in Section \ref{Sec:Conc}.


\section{Observations}
\label{Sec:Obs}

Phaethon made its closest approach to Earth on 16 December 2017 when it came to 0.069 AU of Earth. This was the closest Phaethon has been to Earth since its discovery in 1983 and provided an unprecedented opportunity to observe this apparent asteroid for possible cometary activity. Though outgassing will likely be highest when the asteroid is nearer perhelion, water ice-driven cometary activity can be expected at all distances less than about 3 AU from the Sun \citep{Delsemme71}, and our sensitivity to such activity is much increased when we pass the asteroid closely. The previous closest approaches were 21 December 1984 at 0.24 AU and 10 December 2007 at 0.12 AU\footnote{JPL Solar System Dynamics website https://ssd.jpl.nasa.gov/sbdb.cgi?sstr=3200, retrieved 9 April 2019}.

Using multiple ground-based telescopes, we obtained data from Phaethon as it was approaching the Earth to provide a thorough examination of this interesting body. Nine nights of multicolor photometry with the 0.6 m telescope at the Xingming Observatory provide for high-cadence color information and a detailed phase curve. Gemini North and Canada-France-Hawaii Telescope observations during this period allow an assessment of possible cometary activity and meter-scale fragmentation, and to determine Phaethon's current mass-loss rate, if any. The observations are summarized below.


\subsection{CFHT}
\label{Sec:CFHT}

Observations were conducted using the MegaPrime/MegaCam wide-field imager at the 3.6 m Canada-France-Hawaii Telescope. The charge-coupled device (CCD) detector is a mosaic of thirty-six $2112 \times 4644$ pixel chips, covering a total area of roughly $1 \times 1 ~$deg$^2$ on the sky at 0.187 arcsecond per pixel \citep{Boulade03}. The observations spanned 2017-Nov-15 12:30:22.74 UT -- 13:24:43.16 UT during which we obtained eight 30-second exposure images, six of which were taken with Phaethon being tracked on the sky, the other two were tracked sidereally to allow stars in the field to provide stellar point-spread functions (PSFs) for comparison with Phaethon. These exposures were chosen to search for faint coma or other activity near the asteroid (see Section \ref{Sec:Coma}). We also obtained 4 images with a longer exposure time (900 s) which were used for searching for possible meter-sized fragments of Phaethon comoving with it in the volume of space around it (see Section \ref{Sec:Fragments}). Because of the demands of queue operations at the telescope, CFHT images were taken a few weeks before, and not at, Phaethon's closest approach. Phaethon's heliocentric and geocentric distances were $R = 1.449$ AU and $\varDelta = 0.611$ AU during the time of observations.


\subsection{Gemini}
\label{Sec:Gemini}

We obtained images of Phaethon during 2017-Dec-13 11:34:50.2 UT -- 12:33:04.2 UT using the Gemini Multi-Object Spectrograph on Gemini North (GMOS-N). The GMOS-N Hamamatsu array detector consists of three $2048 \times 4176$ chips of two different types arranged in a row. The images were obtained using the 12-amp mode. The Hamamatsu array detector has a resolution of 0.0807 arcsecond per pixel. Twenty-four images with 2 second exposure times are analyzed here to search for coma (see Section \ref{Sec:Coma}), where in each case the asteroid was on the central chip (chip 6) near the center. Phaethon was tracked in all images except for 8, which were siderally tracked to allow field stars to provide prototype PSFs. Phaethon's heliocentric and geocentric distances were $R = 1.067$ AU and $\varDelta = 0.094$ AU during the time of observations.


\subsection{Xingming Observatory}
\label{Sec:Xingming}

Observations were made using Ningbo bureau of Education and Xinjiang observatory Telescope (NEXT), a 60-cm reflector at Xingming Observatory in Mt. Nanshan, Xinjiang, China. The observatory coordinates are $43{\degr} 28^\prime 15'' N$ and $87{\degr} 10^\prime 39.6'' E$ and it is at an elevation of $2,080 $ m (IAU Code: C42). The FLI PL230 CCD camera used to obtain the data has a resolution of $0.6''$ per pixel. It has an array of $2048 \times 2064$ pixels, a field of view of $22 \times 22$ arcmin, and a focal length of $4800$ mm. We used a total of 5245 light frames obtained between 2017-Dec-11 and 2017-Dec-19 to measure the colors and phase curve (see Section \ref{Sec:Photometry}). Phaethon's heliocentric and geocentric distances changed from $R = 1.109$ to $0.966$ AU and $\varDelta = 0.133$ to $0.085$ AU during the time of observations as it approached the Earth. The frames were taken with Johnson BVRI filters and exposure times of 2, 3, or 6 seconds. 


\section{Results}
\label{Sec:Results}
\subsection{Coma Search: CFHT and Gemini Data}
\label{Sec:Coma}

If Phaethon exhibits no coma, its light profile should be that of a point source such as a star. If the asteroid does have detectable extended emission due to dust or gas, its point-spread function will be broader and it is that broadening we are searching for here. We compared an untrailed image of Phaethon to untrailed images of field stars from the same detector chip. The fluxes were averaged in concentric annuli centered on the centroids of the objects found using the Aperture Photometry Tool (APT)\citep{Laher12}. Any central pixels of Phaethon at or near saturation were removed before fitting. In images where a field star appeared close to Phaethon, which included all CFHT images and 6 of the Gemini images, a wedge was drawn $\pm 30 \degr$ from the center of that star and was excluded before calculating the mean along each annulus. Two Gemini images with multiple background stars near Phaethon were discarded.

A Moffat function of the form shown by Equation \ref{Eq:Moffat} \citep{Trujillo01} was fit to each light profile.

\begin{equation}
\label{Eq:Moffat}
PSF(r) = A \left[1+\left(\frac{r}{\alpha}\right)^2\right]^{-\beta} ~,
\end{equation}

\noindent where $A$ is the amplitude, and $\alpha$ and $\beta$ are related by full-width half maximum (FWHM) through $FWHM = 2\alpha \sqrt{2^{1/\beta}-1}$ with $PSF(FWHM/2) = (1/2)PSF(0)$. When fitting the profile of the star, all three parameters were allowed to vary by the fitting procedure; however, for Phaethon $\beta$ was set to that found by fitting the Moffat fit to the star. The light curves were then normalized to the peaks of the Moffat fits. For images where Phaethon was saturated or close to saturation by $10\%$, we excluded those points from the Moffat fit. This was the case for all Gemini images except for 3. Sky background was subtracted by taking the mean of a few points in the tails of the distributions and subtracting from the data before the fits were made. 

\begin{figure*}
    \centering
    \includegraphics[width=13cm]{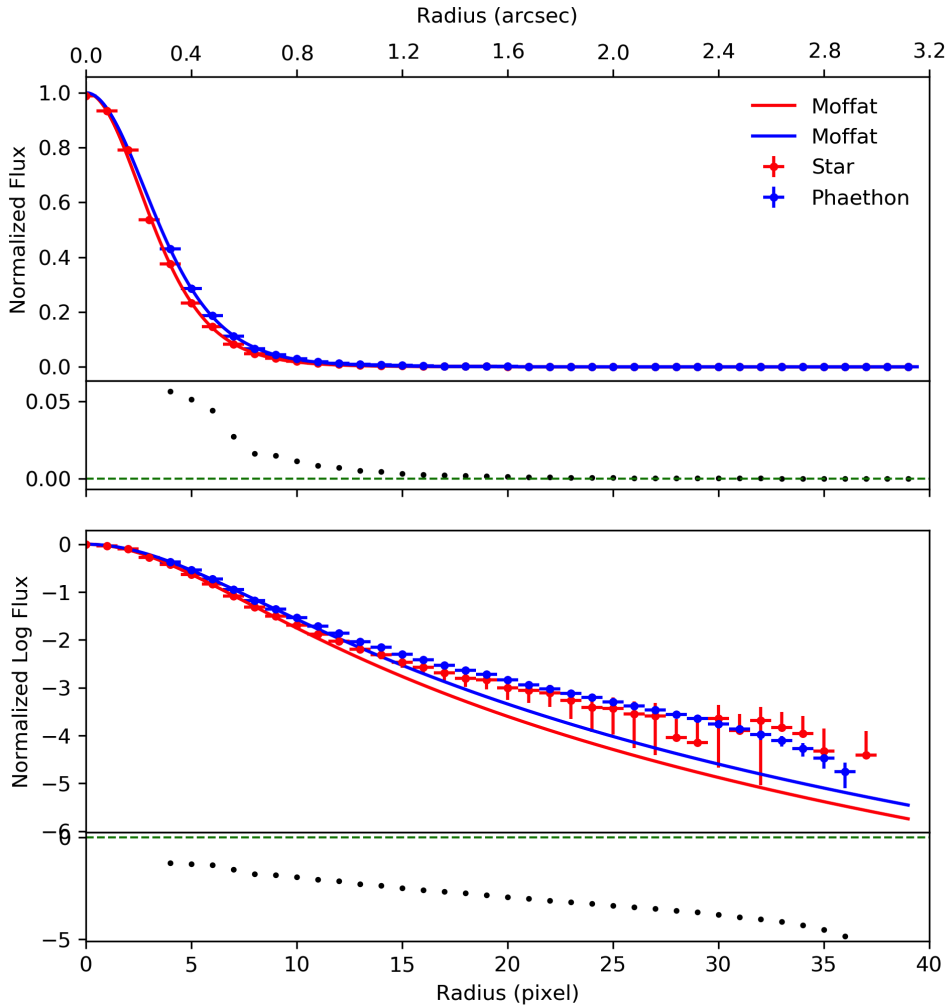}
    \caption{Azimuthally averaged surface brightness distribution of asteroid (3200) Phaethon from one of the Gemini images. The surface brightness distribution of a field star from the same chip is shown for comparison. The light curves have been normalized to the peaks of the Moffat fits. The Moffat fit to the star (red curve) is subtracted from Phaethon's light profile (blue points) and the residuals are shown by the black dots at the bottom of the plot. The profiles are shown in linear (top) and log (bottom) spaces. Error bars are Poissonian, proportional to the square root of the number of remaining photons in each annulus.}
    \label{Fig:PSF_G}
\end{figure*}

The Moffat fits were weighted by the uncertainties, which were assumed Poissonian and proportional to the square root of the number of photons in each annulus. The light curve of Phaethon with the Moffat stellar profile subtracted was in order to set an upper limit on dust production. An example representing the median mass-loss rate measured is shown in Figure \ref{Fig:PSF_G}.

\subsubsection{Upper Limit on Mass-Loss Rate}
\label{Sec:MassLoss}

Although the mechanisms for Phaethon's dust ejection near perihelion are not well understood, attempts have been made to calculate the mass-loss rate of Phaethon from observations to determine if it matches the total mass of the Geminid stream \citep[see for instance,][]{Jewitt13, Hui17, Ye18}. However, to date all calculations for Phaethon's dust production rate have resulted in values that are insufficient to explain the mass of the stream.

The profiles derived for Phaethon do typically show a small excess over the profiles of nearby stars, but are all consistent within the errors with no dust production. Our primary uncertainty is in the location of the center of the PSFs. Though we use a centroiding algorithm which gives sub-pixel accuracy, our data points still have horizontal uncertainties of about 0.5 pixels, or $\sim10$\% of the PSF width. The fact that the residuals are highest where the PSFs are steepest and then drop off (see Figure~\ref{Fig:PSF_G}) is also a hint that uncertainty is dominating the effect we see (The size of the Gemini GMOS pixels at 0.08 arcseconds are relatively large, contributing to this effect).  Nonetheless, from the typical excess that is seen, we can calculate an upper limit to Phaethon's mass-loss rate, $dM/dt$ in kg/s, using Equation \ref{Eq:MassLoss} derived from \cite{Meech86} Equations 3-5, which we rewrite as:

\begin{equation}
\label{Eq:MassLoss}
\log \left(\frac{dM}{dt} \right) = \frac{30.7-m_{zero}}{2.5} - \log\left(\frac{p_v ~t ~t_{exp}}{\rho ~a ~R^2 ~\Delta^2 ~N_e}\right)
\end{equation} \\

\noindent where $m_{zero}$ is the zero-point magnitude specific to each detector (see Section \ref{Sec:Obs}), $p_v=0.1066$ is the assumed geometric albedo of the dust grains released by Phaethon, $t$ and $t_{exp}$ are the diaphragm crossing time and the exposure time in seconds, $\rho = 3000 ~$kg~m$^{-3}$ and $a = 1 ~\mu$m are the grain density and radius \citep{Jewitt13}, $R$ and $\varDelta$ are the heliocentric and geocentric distances in AU obtained from JPL's Horizons system (https://ssd.jpl.nasa.gov/horizons.cgi), and $N_e$ is the number of electrons between Phaethon's and the stellar profile within the aperture.

Using Equation \ref{Eq:MassLoss}, we calculated an upper limit to Phaethon's mass-loss rate in our Gemini observations (geocentric distance $\varDelta = 0.094$ AU, heliocentric distance $R=1.067$ AU) at $0.2 \pm 0.1$ kg/s. From our CFHT observations which were made when Phaethon was further away from the Sun ($R = 1.449$ AU and $\varDelta = 0.611$ AU), we found the mass-loss rate to be at most $0.06 \pm 0.02$~kg/s. Due to their lower sensitivity, we did not use the Xingming observations for calculating the mass-loss rate.

Previous studies indicate that Phaethon shows consistent mass loss only near perihelion ($q=0.14$ AU) at a rate of 3 kg/s \citep{Jewitt13}. Observations of Phaethon in 2003 using the 2.2 m University of Hawaii telescope, \cite{Hsieh05} put an upper limit on its mass-loss rate of $\sim 0.01$ kg/s when Phaethon was at $R = 1.60$ AU and $\varDelta = 1.39$ AU. Our results are consistent with these, and with very low or absent dust production from Phaethon during its close approach to Earth in 2017.


\subsection{Photometry: Xingming Data}
\label{Sec:Photometry}

The set of over 5000 raw CCD images were bias, dark and flat-field corrected, then astrometrically calibrated with astrometry.net \citep{Lang10}. Sources in the images were located using Source Extractor \citep{Bertin96}, and Phaethon identified from its on-sky position from an ephemeris retrieved from JPL's Horizons system\footnote{https://ssd.jpl.nasa.gov/horizons.cgi}. A photometric calibration of each image is then performed by matching the remaining sources with stars in the UCAC4 catalog \citep{Zacharias13}. The median numbers of star matches were 47, 65, 66, and 43 per image in the B,V,R and I filters respectively. The UCAC4 catalog provides Johnson B and V and Sloan r' and i' magnitudes. The r' and i' magnitudes were converted to Johnson R and I magnitudes for comparison with the observations via the equations of \cite{Jordi06}. The raw images were taken in sets of 5 exposures of 2, 3 or 6 seconds per filter, and these were stacked to improve the signal-to-noise. Sixteen observations more than 0.5 magnitudes from the daily median are rejected, as well as five anomalous points at the beginning or end of exposures sets. One two-hour set of data with a lack of suitable I-band calibrators on-image was corrected from data taken before/after; one three-hour stretch of B-filter data on Dec 18 with small numbers of calibrator stars on-image was omitted, leaving 1019 photometric measurements across the four filters. Colors are constructed between stacked images separated by no more than 5 minutes in time (median time between exposures used in color calculations is 1.8 minutes).


\subsubsection{Colors}
\label{Sec:Colors}

The mean colors of Phaethon derived from our observations are $B-V = 0.702 \pm 0.004$, $V-R = 0.309 \pm 0.003$ and $R-I = 0.266 \pm 0.004$ where the accompanying uncertainties represent the standard error of the mean. The asteroid's colors are neutral overall, slightly redder than solar in B-V, and slightly bluer in V-R and R-I, consistent with the classification of Phaethon as a F-type or B-type asteroid \citep{Tholen85, Green95}\footnote{For reference, solar colors are $B-V = 0.67$, $V-R = 0.36$ and $R-I = 0.35$ \citep{Hartmann90, Jewitt01}}.

There are few published colors for Phaethon for comparison, but these are listed in Table~\ref{Tab:colors}. Ours are consistent with all the published values within or near the uncertainties except for our $B-V$ colors which are redder and closer to solar colors than the other published values.
\begin{table}[htbp]
    \centering
    \begin{tabular}{c|ccc}
         ref & B-V & V-R  & R-I \\
        \hline \hline
        \cite{Skiff96}              & -               & 0.34          & -  \\
        \cite{Dundon05}             & $0.59 \pm 0.01$ & $0.35 \pm 0.01$ & $0.32 \pm 0.01$ \\
        Kasuga \& Jewitt (2008)     & $0.61 \pm 0.01$ & $0.34 \pm 0.03$ & $0.27 \pm 0.04$ \\
        Jewitt (2013)               & $0.67 \pm 0.02$ & $0.32 \pm 0.02$ & - \\ 
        Ansdell+ (2014) 1997 Nov 12 & $0.58 \pm 0.01$ & $0.34 \pm 0.01$ & - \\
        Ansdell+ (2014) 1997 Nov 22 & $0.57 \pm 0.01$ & $0.36 \pm 0.01$ & - \\
        Ansdell+ (2014) 1995 Jan 4  & $0.52 \pm 0.01$ & $0.33 \pm 0.01$ & - \\
        Lee (2019)                  & $0.64 \pm 0.02$ & $0.34 \pm 0.02$ & $0.31 \pm 0.03$ \\
        This work                   & $0.702 \pm 0.004$ & $0.309 \pm 0.003$ & $0.266\pm0.004$ \\
        \end{tabular}
    \caption{Published B-V, V-R and R-I colors of Phaethon. \citep{Skiff96, Dundon05, Kasuga08, Jewitt2013b, Ansdell14, Lee19}}
    \label{Tab:colors}
\end{table}

\begin{figure*}
    \centering
    \includegraphics{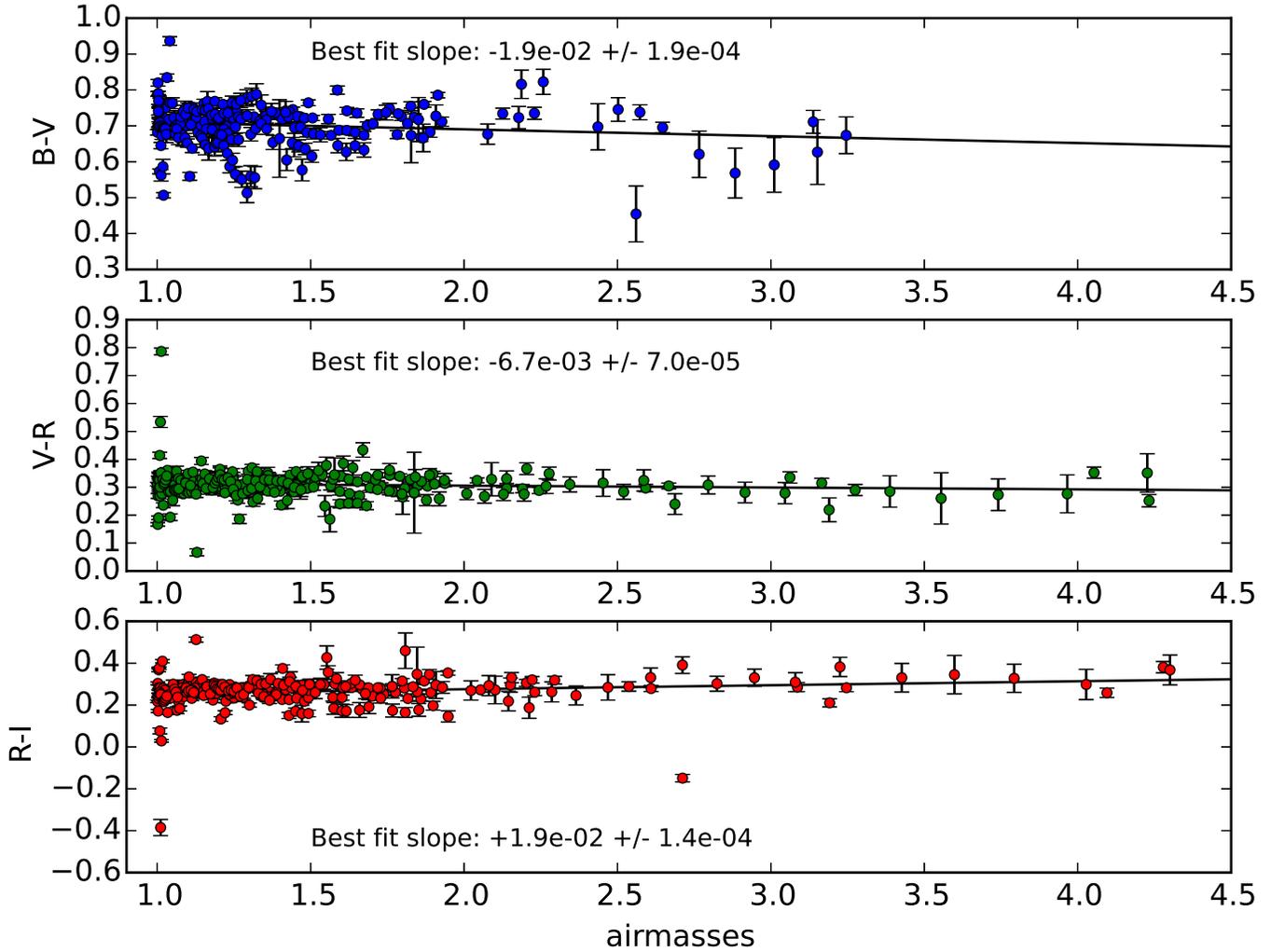}
    \caption{Phaethon's colors as a function of airmass. Best fit slopes weighted by the photometric error are shown.
    }
    \label{Fig:airmass}
\end{figure*}

Could our redder value of B-V be the result of some observational effect? The BVRI filters were cycled through continuously throughout the campaign, and any observations that went into a color determination were at most a few minutes apart. Late in the campaign as Phaethon moved closer to the Sun, the observations were made at larger airmasses, which might be expected to redden the colors. A reddening effect with airmass is actually excluded by our calibration of the asteroid magnitudes against catalog stars visible side-by-side with Phaethon in each image; a plot of color vs airmass is shown in Figure~\ref{Fig:airmass}. Our method shows negligible color changes at airmasses even above 4.

The uncertainties quoted for our colors are the standard errors of the mean, and inherently assume that Phaethon is completely uniform in color and that the dispersion in the measurements is solely due to measurement error. The standard deviations of our colors are 0.06, 0.05, and 0.07 in B-V, V-R and R-I respectively and could be indicative of real surface color variations. However it is difficult to make unambiguous statements since our observations are unresolved disk-integrated colors and Phaethon's pole solution is uncertain. We are attempting to reconstruct a surface map of the albedo and/or colors of Phaethon from the photometric data and will report on this in a later paper, though this effort may not be successful unless an unambiguous pole orientation and detailed shape model becomes available.


\subsubsection{Phase Curve}
\label{Sec:Phase_Curve}

\begin{figure*}
    \centering
    \includegraphics{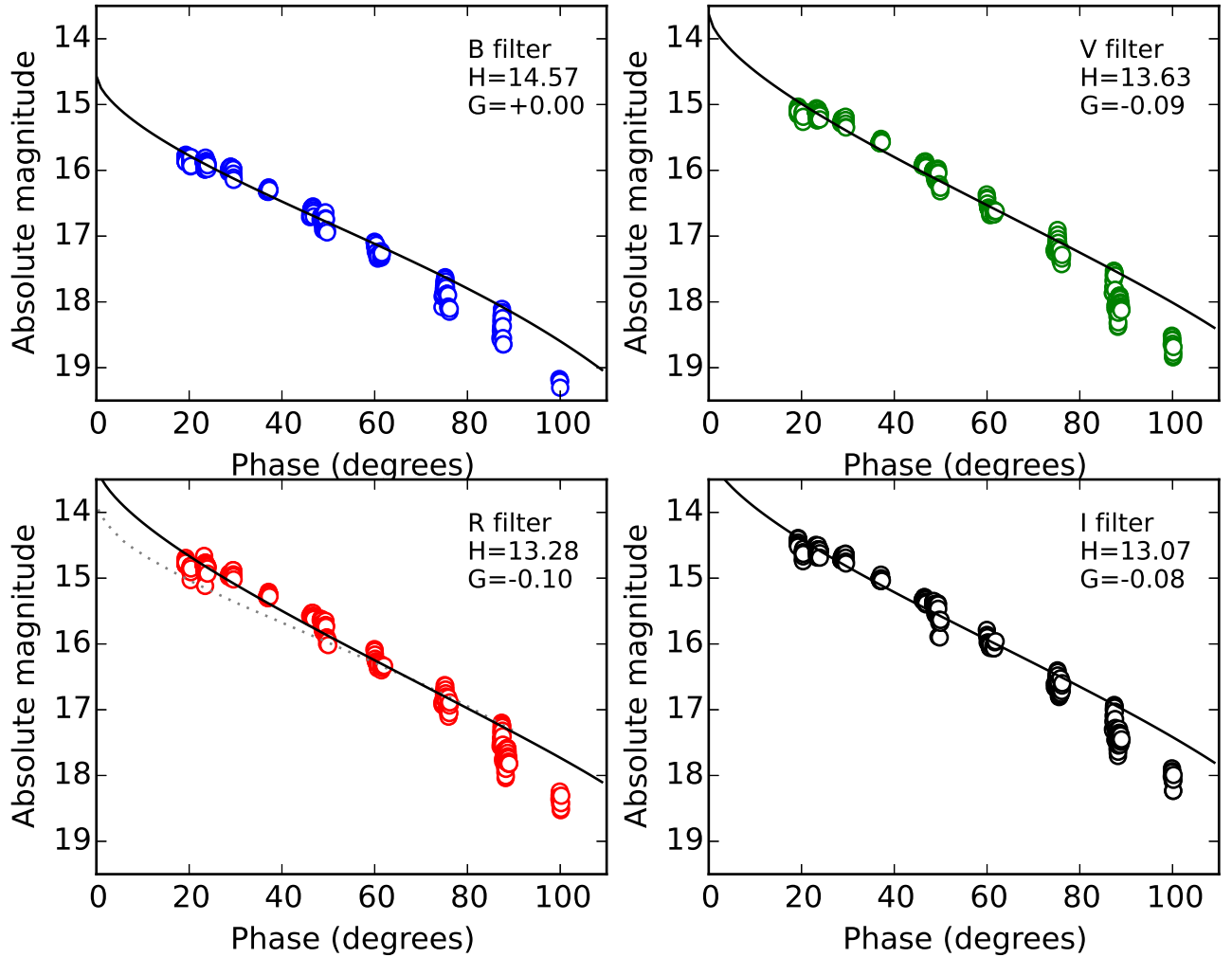}
    \caption{Phase curve of Phaethon in BVRI filters. A fit to the observations using the HG formalism \citep{Bowell89} that minimizes the residuals (weighted for photometric uncertainty) is shown with a black line. The dashed grey line in the bottom left panel shows the fit from \cite{Ansdell14}. The formal error on the observations is smaller than the plotting symbols used.
 }
    \label{Fig:phase_curve}
\end{figure*}

The phase curve of Phaethon is shown in Figure~\ref{Fig:phase_curve}, covering a range of phase angles from 20 to over 100 degrees. The observations are fit using a standard HG function \citep{Bowell89} that minimizes the residuals (weighted by the photometric errors) using Python's scipy.optimize.curve\_fit() \citep{Scipy}; the results are in Table~\ref{Tab:HG}.

\begin{table}[htbp]
    \centering
    \begin{tabular}{c|cc}
         Filter & H  & G \\
        \hline \hline
        B & $14.57 \pm 0.02$ & $0.00 \pm 0.01$  \\
        V & $13.63 \pm 0.02$ & $-0.09 \pm 0.01$  \\
        R & $13.28 \pm 0.02$ & $-0.10 \pm 0.01$  \\
        I & $13.07 \pm 0.02$ & $-0.08 \pm 0.01$  \\
        \end{tabular}
    \caption{Best-fit values to the $H$ and $G$ photometric parameters along with their formal uncertainties.}
    \label{Tab:HG}
\end{table}

The $G$ value in all filters is zero or slightly negative, unusual though permitted within the HG formalism of \citet[see Appendix]{Bowell89}. Negative $G$ values are seen in other F type asteroids e.g. 704 Interamnia \citep{Lagerkvist90}. Our values do differ from some recent results, e.g. \cite{Ansdell14} who found $H=13.90$ and $G=0.06$ in the $R$ band. However, \cite{Ansdell14}'s observations only go up to a phase of $83\degree$ while ours reach $100.2\degree$, and large phase values tend to leverage the HG function down on its right-hand side (e.g. the lower left panel of Figure~\ref{Fig:phase_curve}). Thus we believe our findings don't fundamentally conflict with those of \cite{Ansdell14}. Our data at highest phases (near 100$\degree$) is below our fitted function, but has some of our sparsest coverage. The data covers only 60 minutes (28\% of an asteroid rotation) near the minimum in the rotationally modulated light curve. The light curve observations are presented later in \ref{Fig:light_curve} where the highest phase observations appear in the lower panel to the far right. Thus the lower magnitude measurements at high phase are consistent with rotational modulation of the light curve due to non-sphericity of Phaethon, which is not accounted for in the HG formalism.

\subsubsection{Surface Variations}
\label{Sec:Surf_Var}

\begin{figure*}
    \centering
    \includegraphics[width=8cm]{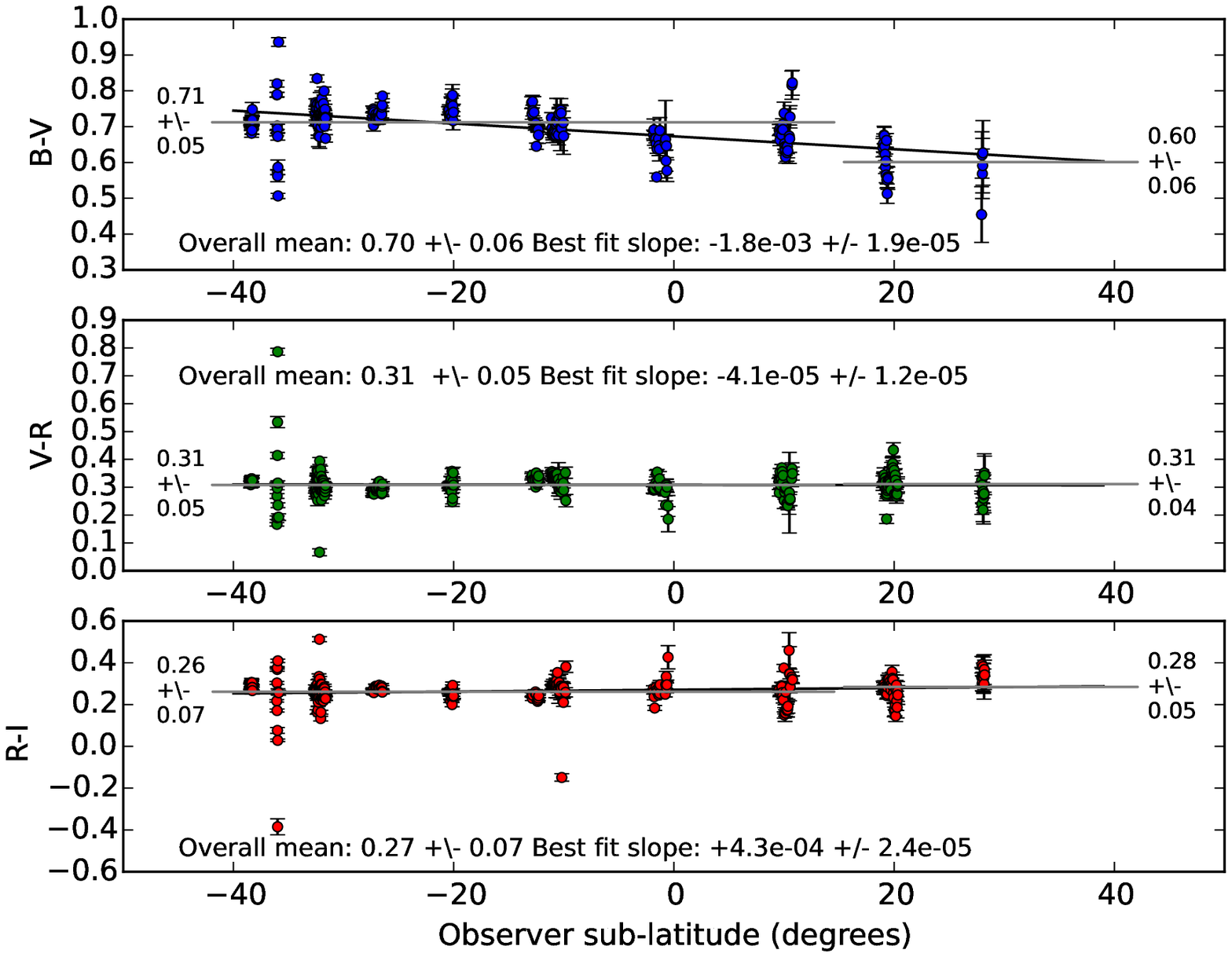}
    \includegraphics[width=8cm]{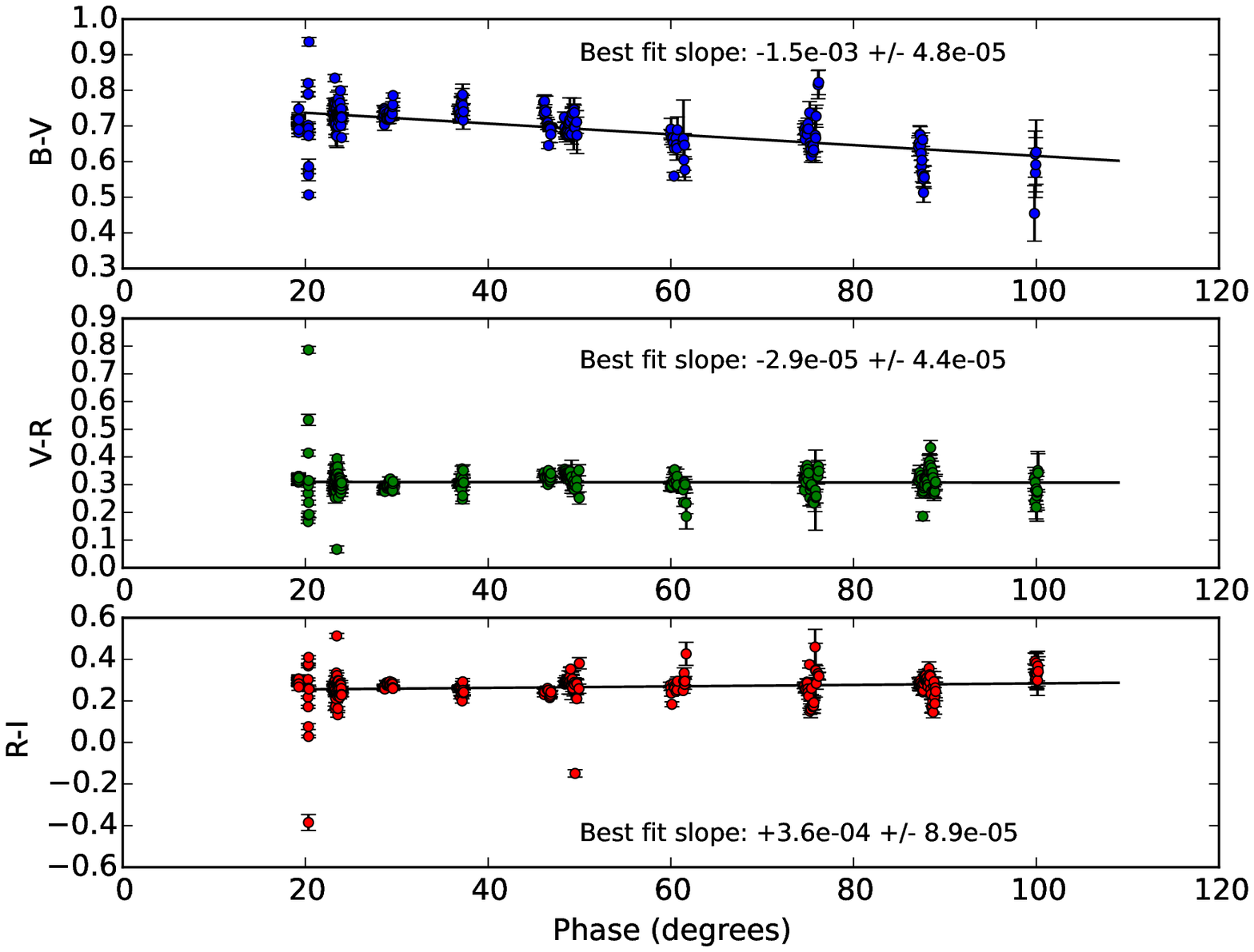}
    \caption{Left panel: Phaethon's B-V, V-R and R-I colors as a function of observer sub-latitude. A linear least-squares fit is shown with a black line. The mean and one standard deviation for the values south and north of a sub-latitude of $+15\degree$ are shown by a grey line. Right panel: Phaethon's B-V, V-R and R-I colors as a function of phase angle. A linear least-squares fit is shown with a black line. 
    }
    \label{Fig:colorsublatphase_curve}
\end{figure*}

The color of Phaethon as a function of observer sub-latitude is shown
in Figure~\ref{Fig:colorsublatphase_curve}.  The pole solution adopted here
is that of \cite{Hanus18}, where the ecliptic longitude of the pole
$\lambda$ is $318.0\degree$, its ecliptic latitude $\beta =
-47\degree$ and the rotation period is 3.603957 hours. We take the
pole whose projection onto the celestial sphere is north of the
ecliptic plane to be its north pole (with positive latitudes)
according to the usual IAU definition \citep{Archinal11}.

\begin{figure*}
    \centering
    \includegraphics[width=8cm,height=6cm]{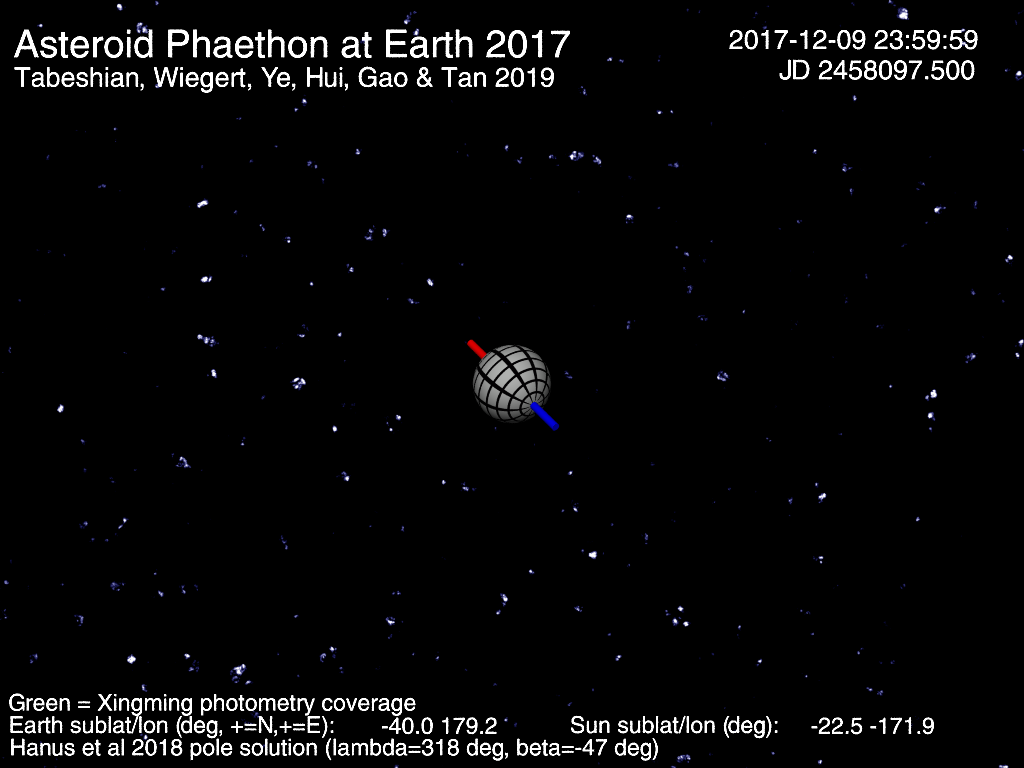}
    \includegraphics[width=8cm,height=6cm]{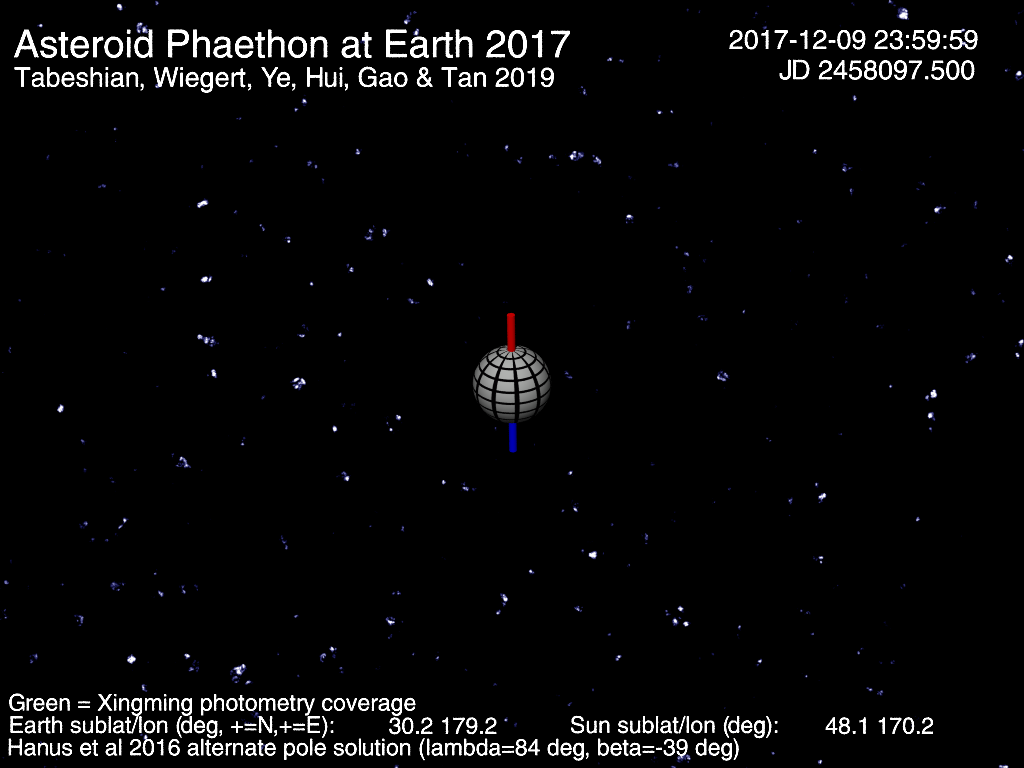}
    \caption{(Animated figure) Viewing geometry from Earth through the close approach of Phaethon in December 2017. On the left is the pole solution adopted here \citep{Hanus18}, on the right is an alternate pole solution from \cite{Hanus16}. Green indicates times where photometric data was taken. Ecliptic north is up.
      {\it These animated figures are permanently available from the Astronomical Journal and semi-permanently at:}\\
      http://www.astro.uwo.ca/$\sim$wiegert/Phaethon/Phaethon-viewinggeom-04.mp4\\
      http://www.astro.uwo.ca/$\sim$wiegert/Phaethon/Phaethon-viewinggeom-05.mp4\\
    }
    \label{Fig:viewinggeometry}
\end{figure*}

Figure~\ref{Fig:colorsublatphase_curve} demonstrates that Phaethon's $B-V$
colors get bluer by 0.1 magnitudes as we move from the southern
hemisphere to the northern, though the $V-R$ and $R-I$ colors do not
show a statistically significant similar trend. The change in $B-V$ is
most apparent in Figure~\ref{Fig:colorsublatphase_curve} at about
+15-20$\degree$ sub-latitude. The mean value of $B-V$ south of this
point is 0.71$\pm 0.05$, while north of it, it is $0.60 \pm
0.06$. Note that the Sun's sub-latitude hardly varies during our
observations but remains between $-22\degree$ and $-24\degree$. As a
result, the illumination of the asteroid remains nearly constant
(except for its rotation) during our observations, while our vantage
point moves from one which sees primarily the asteroid's southern
hemisphere to one looking at its northern hemisphere over time. An
animation of Phaethon's geometry relative to the Sun and Earth during
the observations can be found in the animated
Figure~\ref{Fig:viewinggeometry}.


Because of the monotonic relationship between sub-latitude and the
phase (Sun-Target-Observer) angle in our observations, it is possible
in principle that this effect is really a phase effect: the asteroid
colors as a function of phase are in Figure~\ref{Fig:colorsublatphase_curve}
for reference. If this were the case, the asteroid would simply appear
bluer in $B-V$ (but not $V-R$ or $R-I$) at large phases. 

Though asteroids commonly undergo a reddening of their colors when
observed at higher phases (''phase reddening''), we are not aware of any
other asteroid with a trend to bluer colors with phase, and indeed B
type asteroids like Phaethon have been found not to suffer any color
changes at all with phase \citep{Lantz18}. In addition, laboratory
studies have observed reddening but not blueing as a function of phase
in meteoritic materials \citep{Gradie80, Sanchez12}; and because the
phenomenon is controlled by the absorption coefficient, which is
inversely proportional to the wavelength, the reflected light is
expected to redden with phase on purely microphysical grounds (see
Section 5 of \cite{Sanchez12}).

Our observed color variation in B-V but not in other colors is
consistent with the published spectral data on Phaethon. Comparisons
of the asteroid's spectra show little or no variation in the red part
of the optical spectrum but more pronounced variations in the B band
(blueward of 500nm), which corresponds with what we see in in the
Johnson colors (for example, \cite{Licandro07} Figure 4 and
\cite{Lazzarin19} Figure 3 both show such comparisons).

Our observation of variations in Phaethon's colors with latitude is
nominally in conflict with that of \cite{Lee19} who concluded that
Phaethon ''doesn't have a latitudinal color variation''. However, the
conflict is really rather weak: their data in the northern hemisphere
is limited (they reach a sub-latitude only 0.3$\degree$ north of
Phaethon's equator), while while we see our bluest colors only north
of $+15\degree$. Also \cite{Lee19} do mention a weak decrease
(blueing) in spectral slope as the observer sub-latitude moves north,
an effect which was not statistically significant in their data but
which is in accord with our findings. \cite{Lee19} use \cite{Kim18}'s
pole solution which is similar to the \cite{Hanus18} solution adopted
here.
 
The pole of Phaethon has been determined a number of times in the last
few years and the preferred solutions are clustered near the
\cite{Hanus18} solution ($\lambda = 318\degree$, $\beta=-47\degree$)
adopted here (some examples include $\lambda = 319\degree \pm 5,
\beta=-39\degree \pm 5$ \citep{Hanus16}; $\lambda = 308\degree \pm 10,
\beta=-52\degree \pm 10$ \citep{Kim18}). However, there are alternate
poles possible, typically at similar ecliptic latitude but with the
ecliptic longitude rotated by approximately 125$\degree$, for example,
$\lambda = 84\degree \pm 5, \beta=-39\degree \pm 5$ \citep{Hanus16};
$\lambda = 85\degree \pm 13, \beta=-20\degree \pm 10$
\citep{Ansdell14}. The recent observations of Phaethon by Arecibo are
consistent with both these possibilities \citep{Taylor19}.

If Phaethon were found to have one of these alternate pole solutions,
our observations were in fact of the northern hemisphere first,
transitioning to the southern, but the basic result is
unaffected. This is because the approximate plane containing the
preferred and alternate pole solutions is roughly perpendicular to our
line-of-sight during the asteroid's close approach, and so an
Earth-bound observer's sub-latitude crosses Phaethon's equator at
about the same time in either case (e.g. 13 hours later for the
\cite{Hanus16} alternate pole solution). As a result, our finding of a
color difference with sub-latitude would still stand, but the exact
distribution on the asteroid's surface would be different. A
definitive pole solution is needed to settle the true spatial
distribution of colors on Phaethon's surface.

Given the discussion above, we conclude that our observations reveal
inherent color variations between portions of Phaethon's surface. The
B-V color does not however show a strong rotational modulation, implying large
portions of the asteroid's surface are involved. Though the transition
occurs as the observer moves to more northerly viewing geometries, we
continue to receive reflected light from both hemispheres. The B-V
change occurs as the south pole moves out of view, and so may equally
be the result of a southern red region as a northern blue one. Our
observations and adopted pole solution point to latitudinal color
differences on Phaethon's surface but, because we are seeing
unresolved disk-integrated colors, and Phaethon's pole solution is
uncertain, and we see other color variations occurring on shorter
time scales, we cannot conclude that Phaethon's color variations are
associated purely with its rotational hemispheres. Nonetheless
substantial large-scale color variations are strongly implied by our
observations.

A latitude-dependent color difference on Phaethon could be due to
differences in solar heating. \cite{Ohtsuka09} and \cite{Ansdell14}
found that the northern latitudes of Phaethon would be preferentially
heated and bluer in B-V, using the pole solution of
\cite{Krugly02}. However, \cite{Hanus16} argued, using an updated pole
solution, that Phaethon has not suffered preferential heating over the
last thousands of years.

Comparisons of our results with these previously published ones are
complicated by non-standard definitions of North.  \cite{Ohtsuka09}
``define Phaethon’s north-pole orientation as the sunlit side at the
perihelion'' and \cite{Ansdell14} seem to follow this choice. This
 puts both of their chosen 'north' poles south of the invariable plane, in
contravention of current IAU guidelines \citep{Archinal11}. However,
if we compute the observer sub-latitudes using their \citep{Krugly02}
pole solution and assume this flip of definitions, then our results
qualitatively agree with \cite{Ohtsuka09} and \cite{Ansdell14} that there
is a trend with observer sub-latitude on Phaethon.

The color difference almost certainly stems from compositional
variations across the surface. Determining whether solar heating is
a possible cause will likely require a definitive pole solution but in the
absence of preferential heating, the excavation of fresh material by
cratering events seems a likely cause. Radar observations from Arecibo
\citep{Taylor19} indicate what could be kilometer-sized craters at
latitudes of $+10\degree$ and $+20\degree$ north ('Candidate
Concavities' b and c in their Table 2, also discussed in section
\ref{Sec:Comparison} below). The production of these
craters could have both spread fresh as-yet-un-spaceweathered material
across the asteroid's surface, while releasing a large amount of
material into the Geminid meteoroid stream without the need for
traditional cometary activity.

The Geminid meteoroid stream mass is very uncertain, and has been
estimated to contain from $10^{14}$g \citep{Beech02} to over $10^{16}$
g of material \citep{Hughes89, Blaauw17}, with some calculations
reaching $10^{18}$ g \citep{Ryabova17}. The excavation of a
kilometer-class crater on Phaethon might release $\sim 10^{15}$~g of
material, so the Geminid meteoroid stream could plausibly be populated
by a single or perhaps a few large impacts on its surface, if the
lower estimates of the stream mass are correct. The current rate of
loss of material from Phaethon's surface due to meteoroid impacts (1
ton per year) \citep{Szalay19} is too small to account for the Geminid
stream.


\subsubsection{Comparison with other observations} \label{Sec:Comparison}

\begin{figure*}
    \centering
    \includegraphics{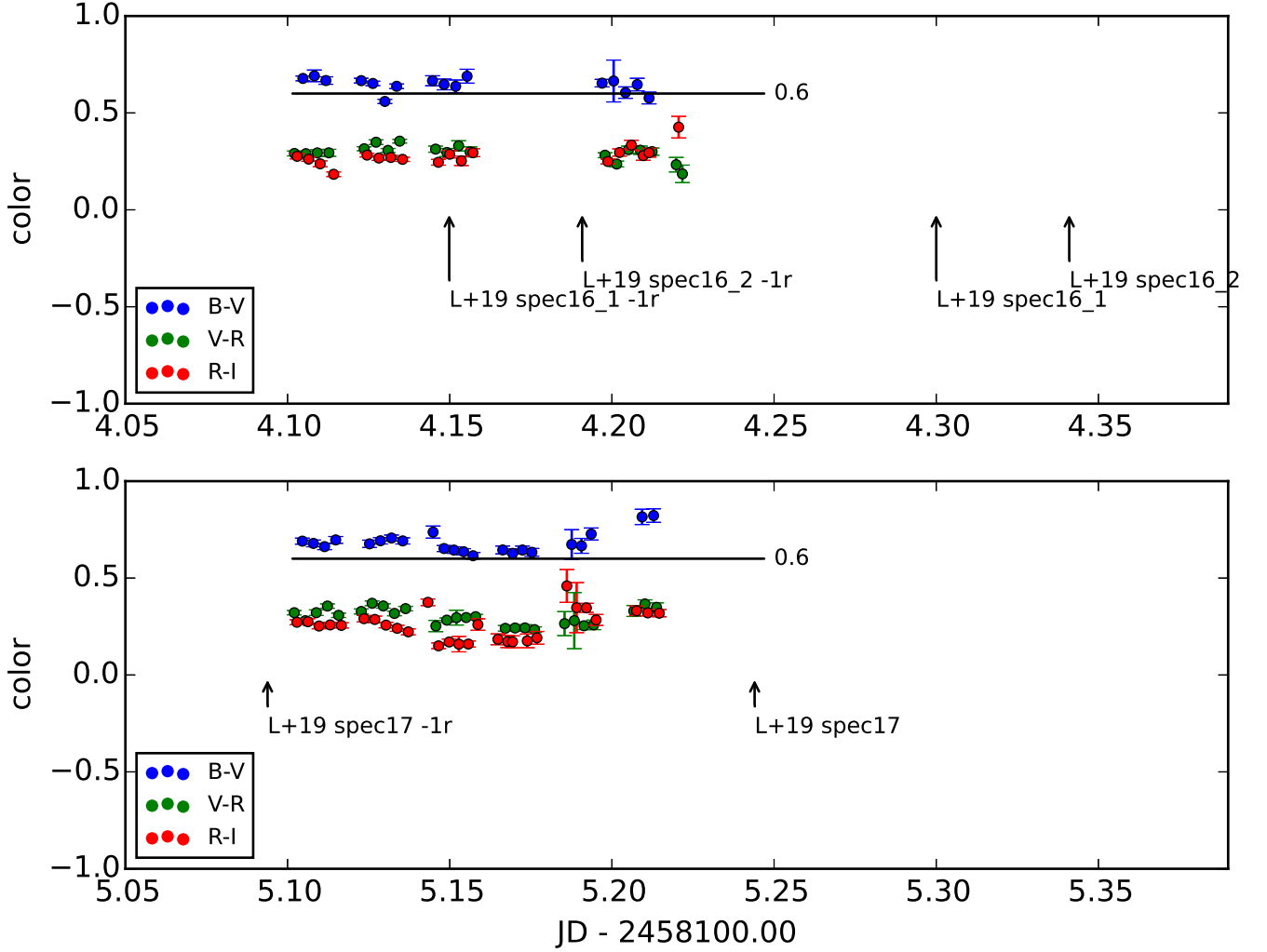}
    \caption{Phaethon's B-V, V-R and R-I colors near the time of \cite{Lazzarin19}'s observations (2 spectra taken Dec 16: spec16\_1 and spec16\_2, and 1 spectrum on Dec 17:spec17). Integer numbers of rotations from their observations are indicated by black arrows e.g. +1r means one rotation later. Our observations that correspond most closely in time with theirs are indicated by the red arrow. We see a reddening of the colors especially in B-V just before their spec17 is taken, which is consistent with their results. 
    }
    \label{Fig:lazzarincolor_curve}
\end{figure*}

\cite{Lazzarin19} report on optical spectra during the close approach, and find a reddening of Phaethon blueward of 500 nm (in the B band) from December 16 to 17. Our observations don't precisely overlap in time with those of \cite{Lazzarin19} but we see something consistent with their reports. Our colors become redder near the end of our observations on Dec 17, just before \cite{Lazzarin19} begin to observe on Dec 17 (see Figure~\ref{Fig:lazzarincolor_curve}). Our observations finish less than an hour before their spectra are taken, corresponding to about a quarter rotation of the asteroid.

\begin{figure*}
    \centering
    \includegraphics{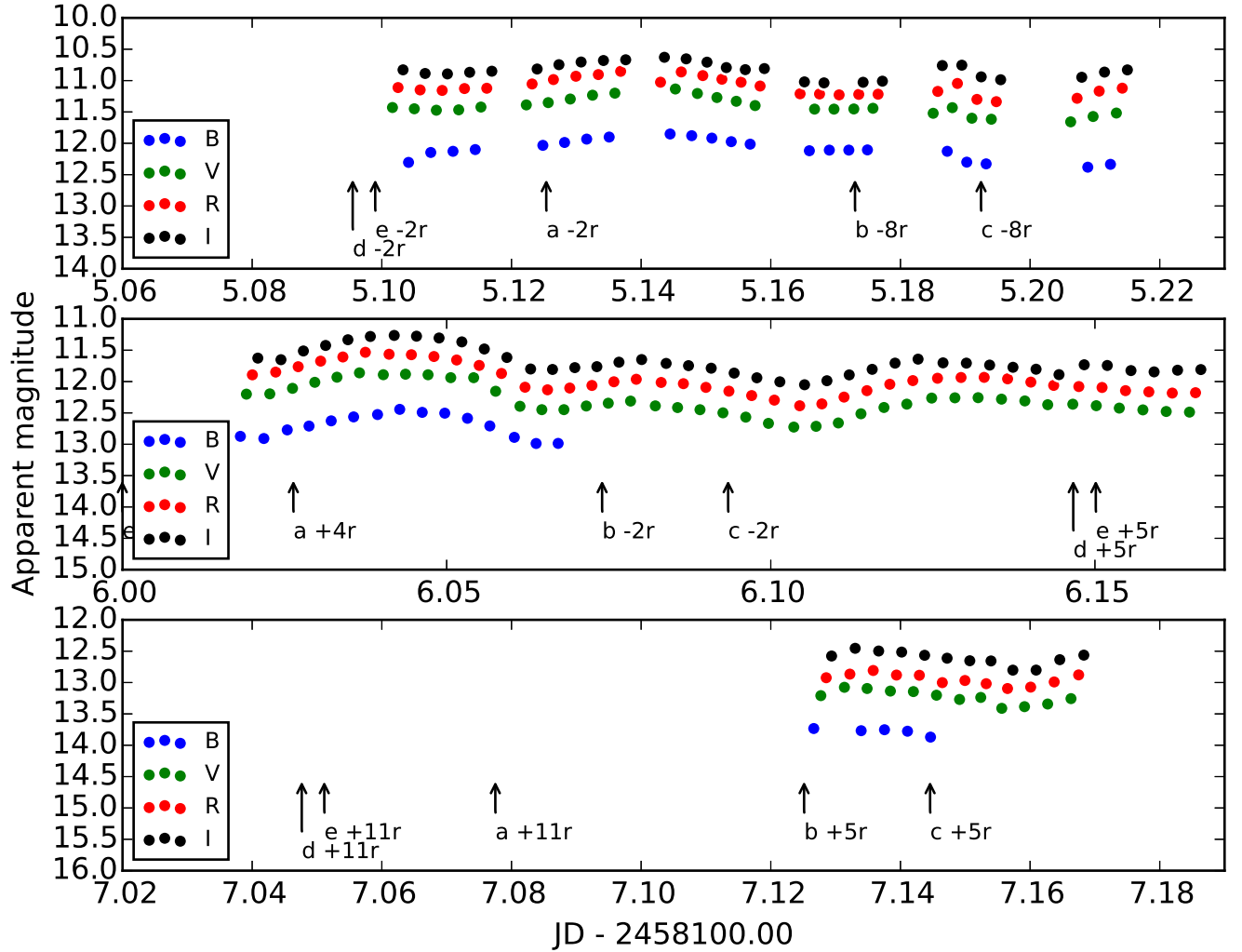}
    \caption{Light curve of Phaethon nearest the features reported by \cite{Taylor19}. Features are: a. Candidate Boulder, b. Candidate Concavity, c. Candidate Concavity, d. Linear Facet, e. Polar Dark Spot. Arrows indicate integer numbers of rotations from the reported sub-observer time e.g. +1r means one rotation later. The formal error bars on the measurements are smaller than the plotting symbols.
 }
    \label{Fig:light_curve}
\end{figure*}

Radar observations of Phaethon by Arecibo \citep{Taylor19} in 2017 indicate surface features (concavities, boulders) that might be correlated with features in the asteroid's light curve or colors. Because the Xingming observatory is on the other side of the globe from Arecibo, our observations are not coincident with theirs. However, we can compare our observations at integer numbers of the asteroid's rotation period, which we here assume is 3.603957 hours \citep{Kim18, Hanus18}. Figure ~\ref{Fig:light_curve} shows our light curve in the BVRI filters during our observations closest in time to the Arecibo observations. The radar-observed features (which \cite{Taylor19} label a through e in their Table 2) are not associated with any obvious changes in the light curve, with the exception of b and c, both termed 'candidate concavities' by \cite{Taylor19}. Both of these occur near broad dips in the light curves in all filters on all three days. Though light curve variations could be accounted for by the non-sphericity of the asteroid, we note that Phaethon's rotation would have carried features at the latitudes of \cite{Taylor19}'s candidate concavities b and c from the far side of the asteroid across the illuminated limb and then across the terminator during this time. The terminator was roughly line with the Earth on the dates of observation (see animated Figure~\ref{Fig:viewinggeometry}), and so any features would have been in sunlight for approximately one hour (0.04 days) on these dates, corresponding to the duration of the dips seen in the light curve. This would provide plenty of opportunity for shadowing effects from obliquely illuminated concavities to create brightness variations. The light curve variations seen in Figure ~\ref{Fig:light_curve} have this time scale, and thus are at least in principle consistent with craters on Phaethon.

Also, returning to the discussion of the phase curves in Section~\ref{Sec:Phase_Curve}, we note that candidate concavities b and c are near the sub-observer point
at the rightmost edge of the lower panel, when our last photometric data was taken and corresponding to our highest phase angles ($\sim 100\deg$). These features correspond
with a relatively dimmer portion of the rotationally modulated light curve (compare, for example, with the middle panel of Figure~\ref{Fig:light_curve}). Thus the relative lowness of our data with respect to the fit at high phase angles in Figure~\ref{Fig:phase_curve} is simply the result of rotational undersampling the light curve at these difficult-to-observe phases.

\subsection{Fragmentation: CFHT Data}
\label{Sec:Fragments}

Given an absence of cometary activity for Phaethon, other mass loss or fragmentation processes must be invoked to explain the Geminid meteor stream. To examine the possibility that Phaethon has undergone coarse fragmentation through collision or other processes, a search for fragments near Phaethon with the Hubble Space Telescope was performed as part of this campaign, but no fragments were seen: these result were reported earlier in \cite{Ye18}. \cite{Jewitt18} performed HST observations of Phaethon near the same time, also without seeing any fragments.

In addition, four CFHT Megacam images (g filter) obtained on 2017-Nov-15 12:30:22.74 UT – 13:24:43.16 UT with 900s exposure times were searched for objects with on-sky motion close to Phaethon's. Tracking at Phaethon's on-sky speed provides improved sensitivity to objects moving along on similar trajectories.

The premise being examined here is that if Phaethon has fragmented in the past, it might still be accompanied by daughter asteroids that are travelling in nearly the same orbit. Such fragments will disperse away from Phaethon over time but if Phaethon underwent such fragmentation recently, then pieces could still be near it.

While at closest approach, Phaethon would have been at an ecliptic latitude of 22 degrees in Andromeda, travelling at a rate of over 2000 arcseconds per hour, making it and any co-moving fragments easy to distinguish from background asteroids. Unfortunately, the geometry at which the images were taken was somewhat before the closest approach owing to the vagaries of the telescope's queue observing system. The images were taken a few weeks early and at that time, Phaethon was at an ecliptic latitude of 13 degrees in the constellation Auriga moving at 20.38 arcsec/hr,  -5.78 arcsec/hr in RA$\cos$(Dec), and 19.55 arcsec/hr in Declination on the sky. Five fragment candidates were identified moving at speeds between 15 and 25 arcsec/hr and with on-sky directions of motion within 5 degrees of Phaethon's during the 0.91 hour arc of the images. All were fainter than apparent g magnitude 22.5 (Phaethon was at $m_g$=16 at the time), and none were identified as known main belt asteroids by the Minor Planet Center's MPChecker\footnote{https://www.minorplanetcenter.net/cgi-bin/checkmp.cgi}.

\begin{figure*}
    \includegraphics[width=8cm,height=8cm]{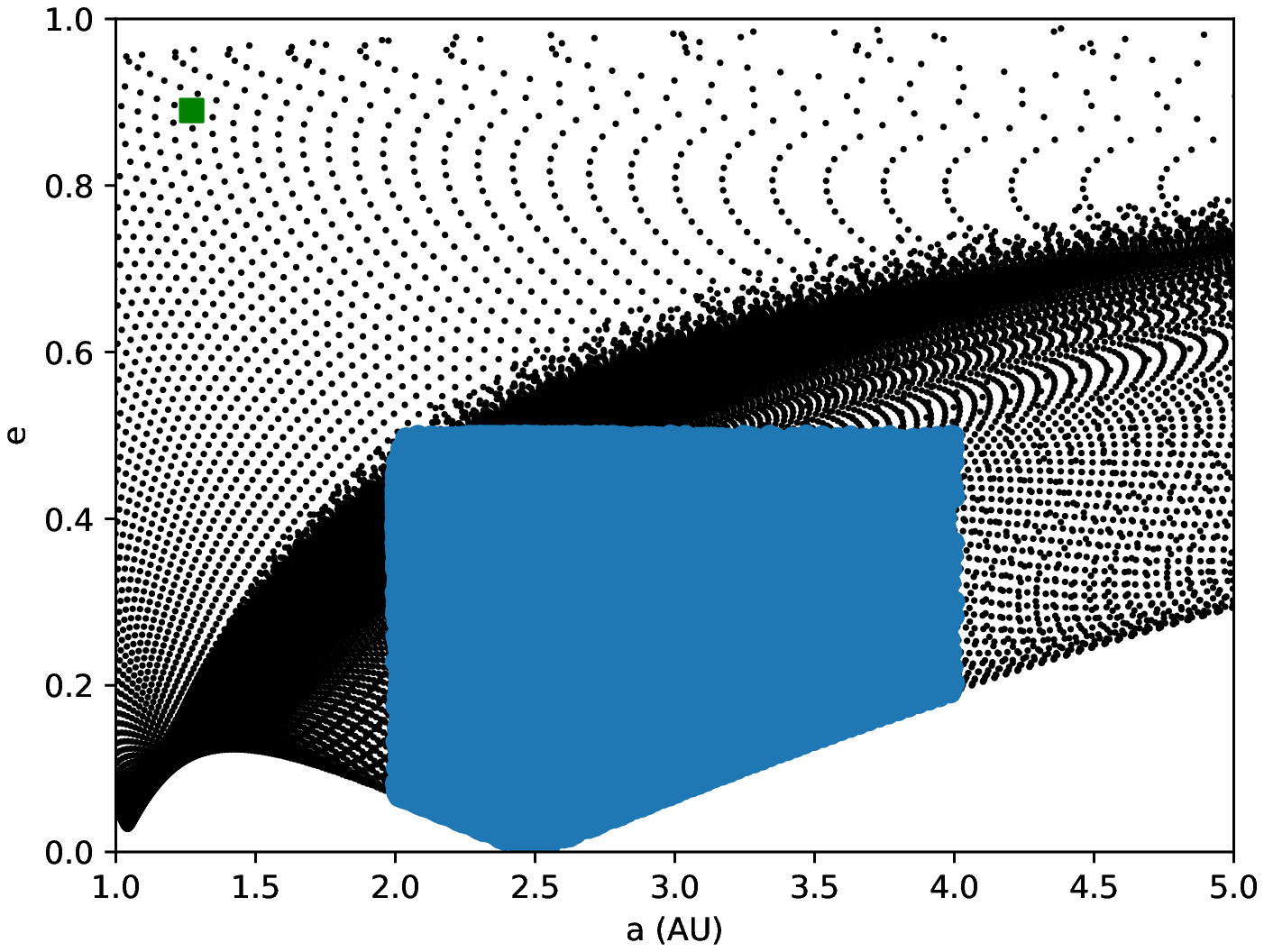}
    \includegraphics[width=8cm,height=8cm]{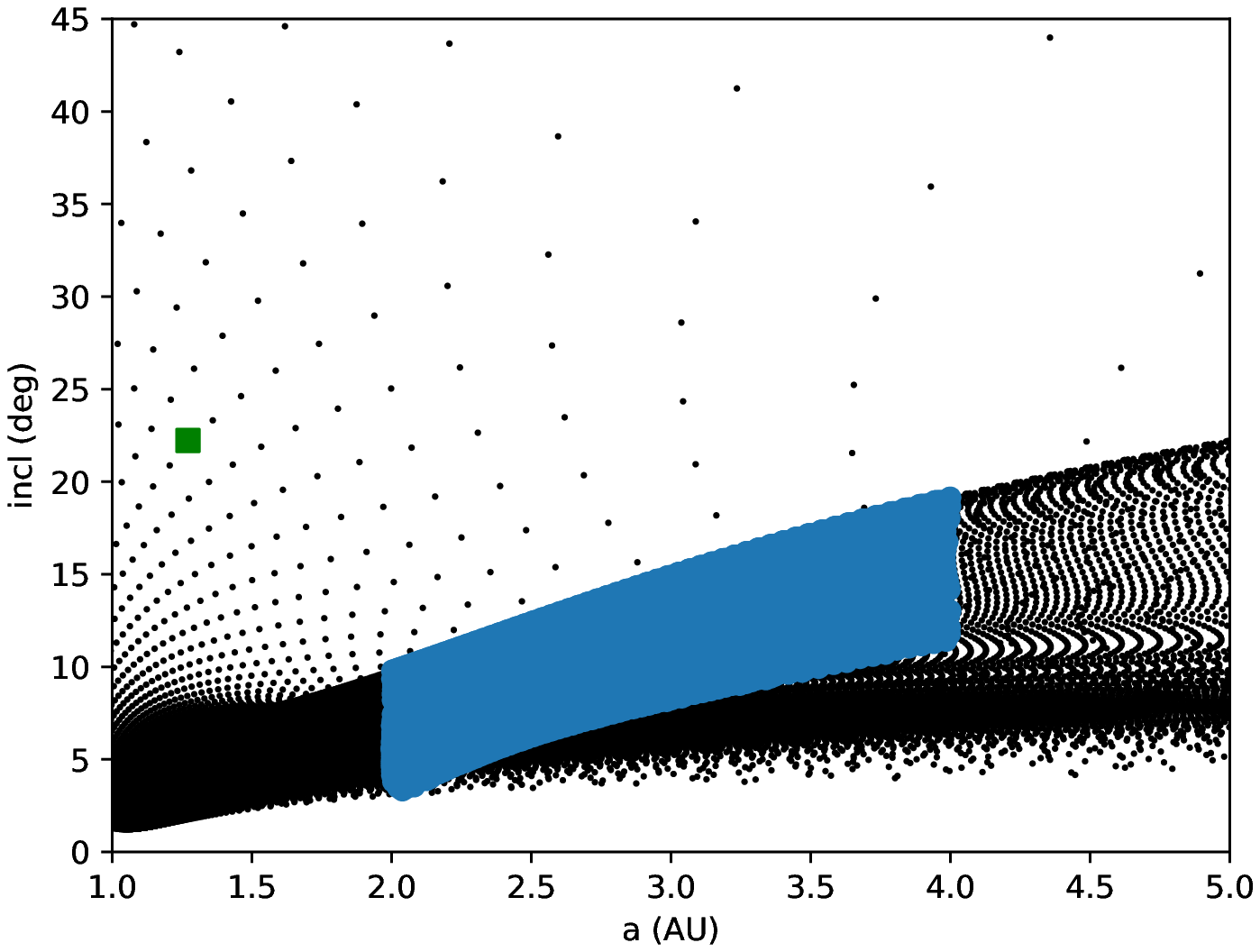}
    \caption{Black dots indicate orbital elements that would have on-sky motions observable from CFHT on 15-Nov-2017 13:00 UT consistent with that observed for Phaethon. Phaethon's values are indicated by the green square. Blue dots indicate values consistent in all three of $a$, $e$ and $i$ with the asteroid belt (see text for details).}
    \label{Fig:onsky}
\end{figure*}

Though the candidate fragments have the on-sky coordinates and rates of motion similar to Phaethon's, it is possible for background or foreground asteroids to do the same but without being on the same orbit as Phaethon. Such objects would have on-sky motions similar to Phaethon's but would not be related to it. We can show that unfortunately, because of the sub-optimal time our observations were taken, main belt asteroids could display rates of motion similar to Phaethon's. Figure~\ref{Fig:onsky} displays the semimajor axis $a$, eccentricity $e$ and inclination $i$ values that could have produced on-sky motions similar to Phaethon's. The blue region indicates orbital elements sets which simultaneously satisfy all three of $2 < a < 4$ ~AU, $e <0.5 $ and $i < 45\degree$, and thus could plausibly be produced by main belt asteroids. Given the abundance of main-belt orbits which could mimic the on-sky motion of Phaethon, we cannot conclude that our candidates are fragments. However, we list them here (and have reported them to the Minor Planet Center) for completeness (see Table \ref{Tab:fragments}), in the hope that a link or absence thereof to Phaethon can be clarified in the future. 
 
\begin{table}[htbp]
    \centering
    \begin{tabular}{ccccc}
        \hline
  Fragment \#  & Date of observation & R.A. & Dec & mag  \\
        \hline \hline
     1 &  2017 11 15.52631 &07 07 33.98 &+36 04 58.9 & 22.8 \\
       &  2017 11 15.53720 &07 07 33.89 &+36 05 02.7 & 22.9 \\
       &  2017 11 15.55315 &07 07 33.76 &+36 05 08.3 & 22.9 \\
       &  2017 11 15.56404 &07 07 33.67 &+36 05 12.2 & 22.6 \\
     2 &  2017 11 15.52631 &07 07 02.07 &+36 10 40.4 & 22.9 \\
       &  2017 11 15.53720 &07 07 01.97 &+36 10 44.2 & 22.7 \\
       &  2017 11 15.55315 &07 07 01.80 &+36 10 50.0 & 22.7 \\
       &  2017 11 15.56404 &07 07 01.69 &+36 10 53.9 & 22.5 \\
     3 &  2017 11 15.52631 &07 07 32.02 &+35 47 28.5 & 24.3 \\
       &  2017 11 15.53720 &07 07 31.90 &+35 47 33.8 & 24.6 \\
       &  2017 11 15.55315 &07 07 31.76 &+35 47 42.1 & 23.1 \\
       &  2017 11 15.56404 &07 07 31.63 &+35 47 46.8 & 24.4 \\
     4 &  2017 11 15.52631 &07 05 30.17 &+35 43 10.2 & 24.7 \\
       &  2017 11 15.53720 &07 05 30.07 &+35 43 15.7 & 24.7 \\
       &  2017 11 15.56404 &07 05 29.84 &+35 43 29.2 & 24.6 \\
     5 &  2017 11 15.53720 &07 08 03.36 &+35 34 42.8 & 23.9 \\
       &  2017 11 15.55315 &07 08 03.23 &+35 34 48.9 & 23.9 \\
       &  2017 11 15.56404 &07 08 03.14 &+35 34 53.3 & 24.3 \\
         \hline
    \end{tabular}
    \caption{Candidate fragment information, including time of observation, Right Ascension, Declination and apparent magnitude in the g filter.}
    \label{Tab:fragments}
\end{table}


\section{Summary and Conclusions}
\label{Sec:Conc}

Ground-based observations made in December 2017. Phaethon's colors and phase curve were measured, and the asteroid was examined for possible cometary activity and fragmentation. Our results are consistent with previous findings of neutral to blue colors overall, though we find a redder B-V color than previous studies. The phase curve of Phaethon is extended to over $100\deg$ in the BVRI filters. Large-scale changes in the B-V colors of Phaethon were observed that are not easily dismissed as phase or airmass effects, while the V-R and R-I colors remain constant. Variations in the light curve consistent with craters reported by \cite{Taylor19} were seen: craters that could be the source of the Geminid meteoroid stream material. An absence of cometary activity down to an upper limit on the mass production rate of $0.06 \pm 0.02$kg/s when the asteroid has at a heliocentric distance of 1.449 AU, and $0.2 \pm 0.1$ kg/s when at a heliocentric distance of 1.067. AU. No fragments were found that could unequivocally be linked to Phaethon.


\acknowledgments

The authors acknowledge the sacred nature of Maunakea and appreciate
the opportunity to observe from the mountain. The Canada-France-Hawaii
Telescope (CFHT) is operated by the National Research Council (NRC) of
Canada, the Institut National des Sciences de l’Univers of the Centre
National de la Recherche Scientifique (CNRS) of France, and the
University of Hawaii. Based on observations obtained at the Gemini
Observatory, which is operated by the Association of Universities for
Research in Astronomy, Inc., under a cooperative agreement with the
NSF on behalf of the Gemini partnership: the National Science
Foundation (United States), National Research Council (Canada),
CONICYT (Chile), Ministerio de Ciencia, Tecnolog\'{i}a e
Innovaci\'{o}n Productiva (Argentina), Minist\'{e}rio da Ci\^{e}ncia,
Tecnologia e Inova\c{c}\~{a}o (Brazil), and Korea Astronomy and Space
Science Institute (Republic of Korea). The operation of Xingming
Observatory and the NEXT telescope was made possible by the generous
support from the Xinjiang Astronomical Observatory and the Ningbo
Bureau of Education. This work was supported in part by the Natural
Sciences and Engineering Research Council of Canada (NSERC, Grant Number
RGPIN-2018-05659) and the University of Western Ontario's
Science and Engineering Review Board Accelerator program. Q.-Z. Ye is
supported by a NASA grant to Thomas Prince. M.-T. Hui is supported by
the Dissertation Year Fellowship at UCLA.


\bibliography{main} 

\end{document}